# Time Series of Magnetic Field Parameters of Merged MDI and HMI Space-Weather Active Region Patches as Potential Tool for Solar Flare Forecasting

PAUL A. KOSOVICH,[1] ALEXANDER G. KOSOVICHEV,[1] VIACHESLAV M. SADYKOV,[2] SPIRIDON KASAPIS,[3]
IRINA N. KITIASHVILI,[3] PATRICK M. O' KEEFE,[4] AATIYA ALI,[2] VINCENT ORIA,[4] SAMUEL GRANOVSKY,[1] CHUN JIE CHONG,[4]
AND GELU M. NITA[1]

[1]Department of Physics, New Jersey Institute of Technology
323 Dr Martin Luther King Jr Blvd, Newark, 07102, NJ, USA
[2]Physics & Astronomy Department, Georgia State University
25 Park Place NE, Atlanta, 30303, GA, USA
[3]NASA Ames Research Center, Moffett Field, 94035, CA, USA
[4]Computer Science Department, New Jersey Institute of Technology
323 Dr Martin Luther King Jr Blvd, Newark, 07102, NJ, USA

## ABSTRACT

Solar flare prediction studies have been recently conducted with the use of Space-Weather MDI (Michelson Doppler Imager onboard Solar and Heliospheric Observatory) Active Region Patches (SMARP) and Space-Weather HMI (Helioseismic and Magnetic Imager onboard Solar Dynamics Observatory) Active Region Patches (SHARP), which are two currently available data products containing magnetic field characteristics of solar active regions. The present work is an effort to combine them into one data product, and perform some initial statistical analyses in order to further expand their application in space weather forecasting. The combined data are derived by filtering, rescaling, and merging the SMARP with SHARP parameters, which can then be spatially reduced to create uniform multivariate time series. The resulting combined MDI-HMI dataset currently spans the period between April 4, 1996, and December 13, 2022, and may be extended to a more recent date. This provides an opportunity to correlate and compare it with other space weather time series, such as the daily solar flare index or the statistical properties of the soft X-ray flux measured by the Geostationary Operational Environmental Satellites (GOES). Time-lagged cross-correlation indicates that a relationship may exist, where some magnetic field properties of active regions lead the flare index in time. Applying the rolling window technique makes it possible to see how this leader-follower dynamic varies with time. Preliminary results indicate that areas of high correlation generally correspond to increased flare activity during the peak solar cycle.

## 1. INTRODUCTION

The study of the magnetic field properties in solar active regions (ARs) helps to increase our understanding of the magnetic activity of the Sun, from global solar cycle-long trends and processes to fast, minute-scale energy releases. The latter is of special importance due to its space weather implications. The free magnetic energy stored in the non-potential fields of ARs can be transformed into radiation in a wide range of wavelengths, the kinetic energy of the coronal mass ejections, and the energy of sub-relativistic solar energetic particles (Aschwanden et al. 2017), which all

Corresponding author: Paul A. Kosovich
paul.kosovich@njit.edu

have different implications for the terrestrial environment (Buzulukova & Tsurutani 2022). This, therefore, makes the studies of the magnetic field properties of ARs important for advancing both fundamental and applied solar physics.

Space-based observations of solar photospheric magnetic fields with the Michelson Doppler Imager onboard Solar and Heliospheric Observatory (SOHO/MDI, Scherrer et al. 1995) and the Helioseismic and Magnetic Imager onboard Solar Dynamics Observatory (SDO/HMI, Scherrer et al. 2012) enabled the continuous monitoring of the line-of-sight (LoS) magnetic fields, with almost unchanged over time quality of observations for each instrument. Space-Weather MDI Active Region Patches (SMARP) and Space-Weather HMI Active Region Patches (SHARP) are two data products derived from these observations, which include maps and summary parameters of automatically tracked solar ARs. These parameters are used as metadata to query the database, and are frequently referred to as "keywords" in primary sources (Bobra et al. 2014, 2021). These data products are developed, and maintained at the SDO Joint Science Operations Center (JSOC), Stanford University. The specific data series utilized in this work, *mdi.smarp_cea_96m* and *hmi.sharp_cea_720s*, are derived from photospheric magnetic field maps in the Lambert cylindrical equal-area coordinates, at 96-minute and 12-minute cadences, respectively. SMARP keywords are based on data from SOHO/MDI and cover the period from April 23, 1996 to October 27, 2010. SHARP keywords are based on data from SDO/HMI and cover the period starting May 1, 2010 onward. Together, these data products span more than two Solar Cycles.

In the context of forecasting space weather events, numerous works have relied on the SMARP and SHARP keywords, or their derivatives. Bobra & Couvidat (2015) were among the first to study the forecasting of M- and X-class solar flares using machine learning, with support vector machines (SVMs) as classifiers. Thirteen SHARP keywords were chosen as input parameters, for the observational period from 2010 to 2014, corresponding to some 1.5 million active region patches. Relatively high predictive scores were obtained, with a True Skill Score (TSS) of $0.761 \pm 0.039$, and a Heidke Skills Score $HSS_2$ of $0.636 \pm 0.037$ in the operational regime. Similar TSS scores were obtained by Bobra & Ilonidis (2016) when using an SVM-based algorithm, with SHARP keywords and Geostationary Operational Environmental Satellite (GOES) X-ray flux as input features, to forecast CMEs produced by M- or X-class flares. SHARP data were especially extensively used for solar flare forecasting (e.g., Liu et al. 2019; Chen et al. 2019; Wang et al. 2020, and many others). Recently, both SHARP (e.g., Sadykov et al. 2021; Abduallah et al. 2022) and SMARP (Kasapis et al. 2022) data products were used separately to forecast Solar Energetic Particle events (SEPs).

One of the challenges in studies of AR properties and their relations to solar transient events is the homogeneity of the observations. Although the number of research attempts trying to predict solar transients is growing, these are typically based on the SDO/HMI vector magnetograms, and limited to Solar Cycle 24, with the exception of some works (e.g., Falconer et al. 2011; Huang et al. 2018; Sun et al. 2022, etc.). In relation to space weather, it becomes clear that, in order to construct a validated forecast and argue about its applicability to operations, one has to consider conditions that are not limited to the Solar Cycle 24. As an example, the recent work by Ali et al. (2023) demonstrated, although based on GOES observations, that the SEP forecasting model trained on the Solar Cycle 24 data demonstrates lower performance with respect to models trained on either of the preceding solar cycles. In such cases, the usage of the LoS magnetic field data spanning two solar cycles becomes an appropriate necessity.

The overlap in high-quality MDI and HMI observations from May 1, 2010 to October 28, 2010, provides an opportunity to re-scale and merge the SHARP and SMARP data products. The motivation for this work is, by using the overlapping observations, to cross-calibrate these data products and create one continuous data series that can be used with machine learning tools and for statistical analysis, with the primary focus on forecasting space weather events, including solar flares, coronal mass ejections (CMEs), and solar energetic particle events (SEPs). The paper is structured as follows: Section 2 describes the data cross-calibration, which includes the data preparation, reduction, and rescaling steps. Section 3 describes the preliminary analysis of the time-lag correlations of the derived homogeneous descriptors with the daily flare index and soft X-ray flux measured by GOES. The results of the analysis are discussed in Section 4 and are followed by conclusions in Section 5.

## 2. DATA PREPARATION

The SHARP and SMARP parameters (keywords) used in the present work are summarized in Table 1. Prior to merging the keywords, rescaling is applied to the following, which are also the primary focus of this study: USFLUXL, MEANGBL, R_VALUE, CMASKL, MEANGBZ, and USFLUXZ. It is also worth noting, that all of the keywords listed in Table 1 are present in both SHARP and SMARP data series, with the exception of UNIX_TIME, MEANGBZ, and USFLUXZ, which are derived here. For a more detailed description of SHARP and SMARP keywords and

**Table 1.** Summary of SHARP and SMARP parameters (keywords) used in the present work

| Keyword | Description | Units | Data Type |
|---|---|---|---|
| T_OBS | Nominal date and time of the observation, TAI | hr:min:sec | date/time |
| UNIX_TIME* | T_OBS expressed as time elapsed since Unix epoch | days | double |
| ARPNUM[1] | The number of the active region patch | number | int |
| NOAA_AR | NOAA active region number that best matches this ARPNUM[1] | number | int |
| NOAA_ARS | List of all NOAA active regions matching this ARPNUM | number | varchar |
| CAR_ROT | Carrington rotation number of CRLN_OBS | number | int |
| USFLUXL[†] | Total line-of-sight unsigned magnetic flux | Maxwells | double |
| R_VALUE[†2] | Unsigned flux near polarity inversion lines | Maxwells | double |
| MEANGBL[†3] | Mean value of the line-of-sight field gradient | Gauss/Mm | double |
| USFLUXZ[*†4] | Vertical component of the total unsigned flux | Maxwells | double |
| MEANGBZ[*†5] | Mean value of the vertical field gradient | Gauss/Mm | double |
| CMASKL[†] | CEA (cylindrical equal-area) pixels in the active region | number | double |
| LAT_FWT | Stonyhurst latitude of the flux-weighted center of active pixels | degree | float |
| CRLT_OBS | Carrington latitude of the observer | degree | float |
| LON_FWT | Stonyhurst longitude of the flux-weighted center of active pixels | degree | float |
| CRLN_OBS | Carrington longitude of the observer | degree | float |
| CDELT1 | Map scale in X direction | degree/pixel | float |
| DSUN_OBS | Distance from SoHO/SDO to center of the Sun | meters | double |
| RSUN_OBS | Observed angular radius of the Sun | arcsec | double |
| QUALITY | Quality index | none | int/hex |

* denotes keywords that are derived in this work, and are not present in the original data series.
† denotes keywords whose SMARP values are here re-scaled, prior to merging with SHARP.
[1] This is called "TARPNUM" in SMARP, and "HARPNUM" in SHARP; renamed here for consistency.
[2] Recomputed here as the common antilogarithm of the R_VALUE.
[3] Units converted from Gauss/pixel to Gauss/Mm.
[4,5] See eq 1 and 2 for derivation.

corresponding formulae, the reader is referred to Table 1 in Bobra et al. (2014), Table 1 in Bobra & Couvidat (2015), and Tables 3 and A.8 in Bobra et al. (2021).

## 2.1. *Data Processing*

The data are retrieved from the Joint Science Operations Center (JSOC) database using the Lookdata tool, accessed through the JSOC web portal[1]. The initial processing step is to convert the JSOC tab-delimited text files into UTF-8 .csv format. Then, the HMI SHARP .csv keyword data files are imported from a specified source directory, along with the MDI SMARP .csv file, any duplicate records from both data series are removed, and SMARP MEANGBL units are converted from Gauss/pixel to Gauss/Mm. For both SHARP and SMARP, the R-value is recomputed by finding the common antilogarithm of the original R-value. Additional filtering is applied to exclude any records corresponding to the Stonyhurst coordinates below -65 and above +65 degrees longitude, in order to avoid the effect of observational distortion near the limb. Low-quality observables are further filtered out based on the QUALITY keyword, which is discussed in Bobra et al. (2021). SHARP records corresponding to QUALITY $\geq$ 65,536 (10,000 in hexadecimal) and SMARP records corresponding to QUALITY $\geq$ 262,144 (40,000 in hexadecimal) represent unreliable Stokes vectors and are therefore excluded. This can also be done at the query stage, using the DRMS Python package (Bobra et al. 2021).

The vertical components of the unsigned flux ($\Phi_z$) and the mean magnetic field gradient ($\overline{|\nabla B_z|}$) are absent in the SMARP data series, because the MDI instrument did not observe the transverse magnetic field components. Therefore, for both SHARP and SMARP, they are derived from the line-of-sight (LoS) parameters as follows:

$$\Phi_z = \frac{\Phi_{LoS}}{\cos(\beta_0) \cdot \cos(\varphi_s) \cdot \cos(\lambda_s) + \sin(\beta_0) \cdot \sin(\varphi_s)} \tag{1}$$

$$\overline{|\nabla B_z|} = \frac{\overline{|\nabla B_{LoS}|}}{\cos(\beta_0) \cdot \cos(\varphi_s) \cdot \cos(\lambda_s) + \sin(\beta_0) \cdot \sin(\varphi_s)} \tag{2}$$

where $\varphi_s$ and $\lambda_s$ are, respectively, the Stonyhurst latitude and longitude of the flux-weighted center of the active pixels, and $\beta_0$ is the Carrington latitude of the solar disk center relative to the observer. We use equations (1) and (2) as an approximation, although the exact relationship between the LOS and true radial values may be more complex (Leka et al. 2017).

In the second stage, the six aforementioned SMARP keywords are rescaled using predetermined parameters (Section 2.3), and the SMARP keywords are merged with their corresponding SHARP keywords at May 1, 2010, 00:00:00, which is the earliest date for which *hmi.sharp_cea_720s* records are available. As a third and final step, all records are grouped by time stamp, data reduction, and integration are performed (Section 2.2), gaps in observations are represented using NaN values, and the output is exported in a comma-separated format.

## 2.2. *Data Reduction and Integration*

The SHARP and SMARP parameters are not one continuous time series, but a collection of time series (one for each solar active region). In the present work, they are converted into true two-dimensional arrays using two different methods, which are plotted for comparison in Figure 1. The first method involves selecting time slices corresponding to R-value maxima, regardless of their active region number; that is, for each time stamp, the record with the highest R-value is chosen regardless of its location on the Sun, and only the keyword values corresponding to that record are included in the reduced time series. R-value is the total weighted unsigned flux density within ~15 Mm of strong-field, high-gradient polarity separation lines, and has been associated with a high probability of initiation of M- and X-class flares (Schrijver 2007). The second method involves integrating the keywords over the entire solar disc: for each time stamp, the keyword values are summed, regardless of their coordinates or active region number. With the latter method, the $\pm 65°$ limb truncation is not applied to the keywords, in order to match them with GOES-15 soft X-ray data, which are not coordinate-specific.

## 2.3. *Determination of Rescaling Parameters*

The fusion of SHARP and SMARP parameters was previously attempted using basic Z-score standardization (Sun et al. 2022), though here we present a novel approach. Ten prominent ARs in the SHARP-SMARP overlap period are selected, with the following NOAA numbers: 11087, 11089, 11092, 11093, 11101, 11106, 11108, 11109, 11112, and 11113. The criteria for their selection are: the peak unsigned flux, the sum of the unsigned flux, the duration of observation, and data quality. Because the SHARP-SMARP overlap period occurred during a solar minimum, a total of 54 NOAA ARs were observed during that time by HMI, out of which only 15 have transited the desired -65 to +65 Stonyhurst longitude range. We exclude two (11073 and 11097) due to incomplete MDI data, and one (11100) due to a large number of zero R-values. We exclude two more (11115 and 11084) because they rank fairly low in terms of both peak USFLUXL, and the sum USFLUXL. It is possible, by changing the selection criteria, to arrive at a different set of ARs, and obtain somewhat different rescaling parameters, although this is outside of the scope of the present work. Since the focus here is on larger active regions that are more or less completely observed, incorporating data from many smaller or partly visible ARs could introduce a degree of uncertainty that is difficult to quantify. Conversely, the inclusion of one or two additional ARs should not have a significant effect on the rescaling parameters, or on the final statistical analysis (Section 3).

For the ten selected ARs, correlating the data involves time-matching each SMARP record with the nearest SHARP record within a $\pm 13$ minute sensitivity window. If no such record is found, the data point is dropped. The remaining time-matched SHARP and SMARP values are then plotted against each other (Figure 2), and linear regression is performed using Ordinary Least Squares (OLS) and Total Least Squares (TLS) methods, of which only the latter is used for rescaling in the present work. The procedure for calculating the Standard Deviation of the Slope (SDS) for TLS regression is based on the work by Tellinghuisen (2020), where it is covered in some detail. The Standard Error (RSE) for TLS is computed here as follows:

$$RSE_{TLS} = \sqrt{\frac{\sum_{i=1}^{n}(y_i - b - mx_i)^2}{(n-2)(1+m^2)}}$$

(3)

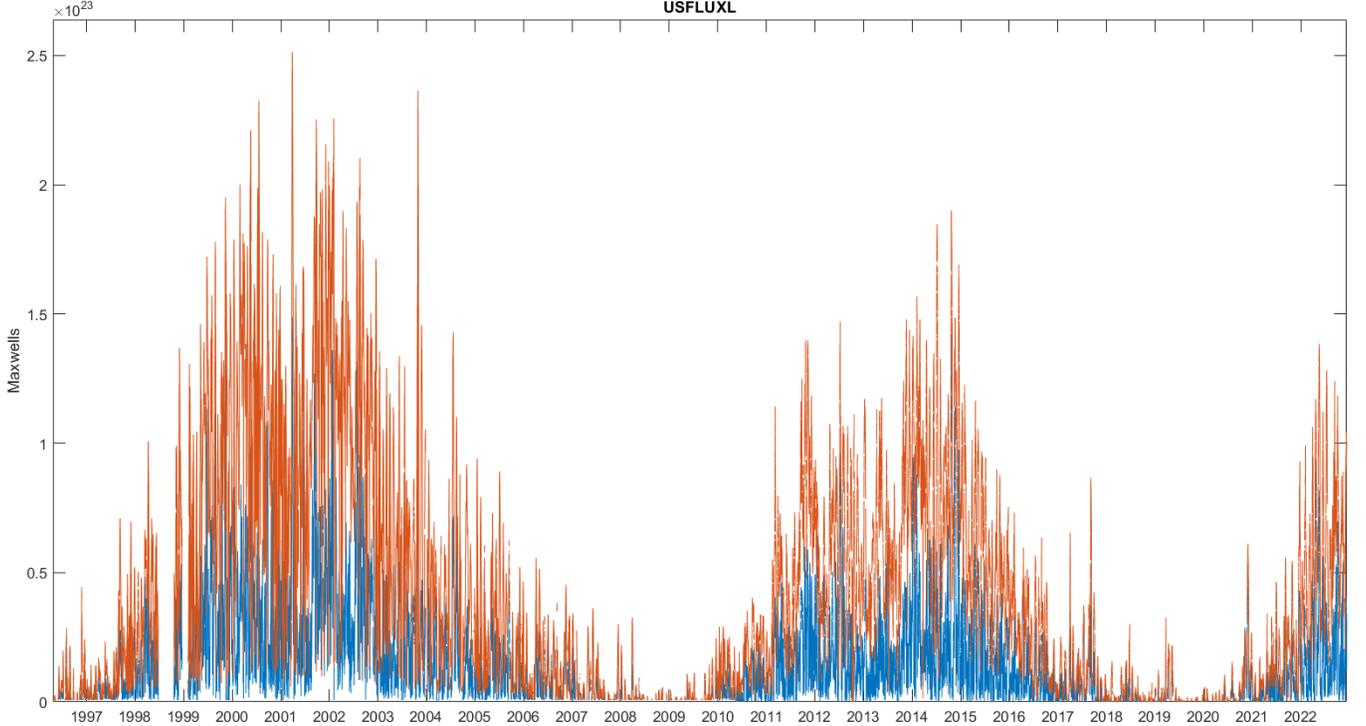

**Figure 1.** The rescaled and merged SMARP and SHARP line-of-sight unsigned magnetic flux (USFLUXL) time series, produced by selecting records corresponding to R-value maxima observed between $-65°$ and $+65°$ Stonyhurst longitude (blue), and by integrating the parameter (in this example, USFLUXL) over the entire solar disk (red). Note the gap in data coverage between June and October of 1998, reflecting the loss and recovery of the SOHO spacecraft (Vandenbussche 2006). There are numerous smaller data gaps, which may not be readily apparent at this scale.

Finally, the rescaling parameters (slope and intercept) for each of the ten ARs are tabulated, and their weighted averages are determined for each keyword, using equation 4.17 in Bevington & Robinson (2003):

$$\mu' = \sum\left(\frac{x_i}{\sigma_i^2}\right)\left[\sum\left(\frac{1}{\sigma_i^2}\right)\right]^{-1}$$

(4)

A brief remark on notation. In equation (3), $x_i$ and $y_i$ are the coordinates of each data point (as in Figure 2b), and $m$ and $b$ are the regression slope and its y-intercept, respectively. In equation (4), $x_i$ is the value of each fitted parameter, and $\sigma_i$ is its statistical standard deviation.

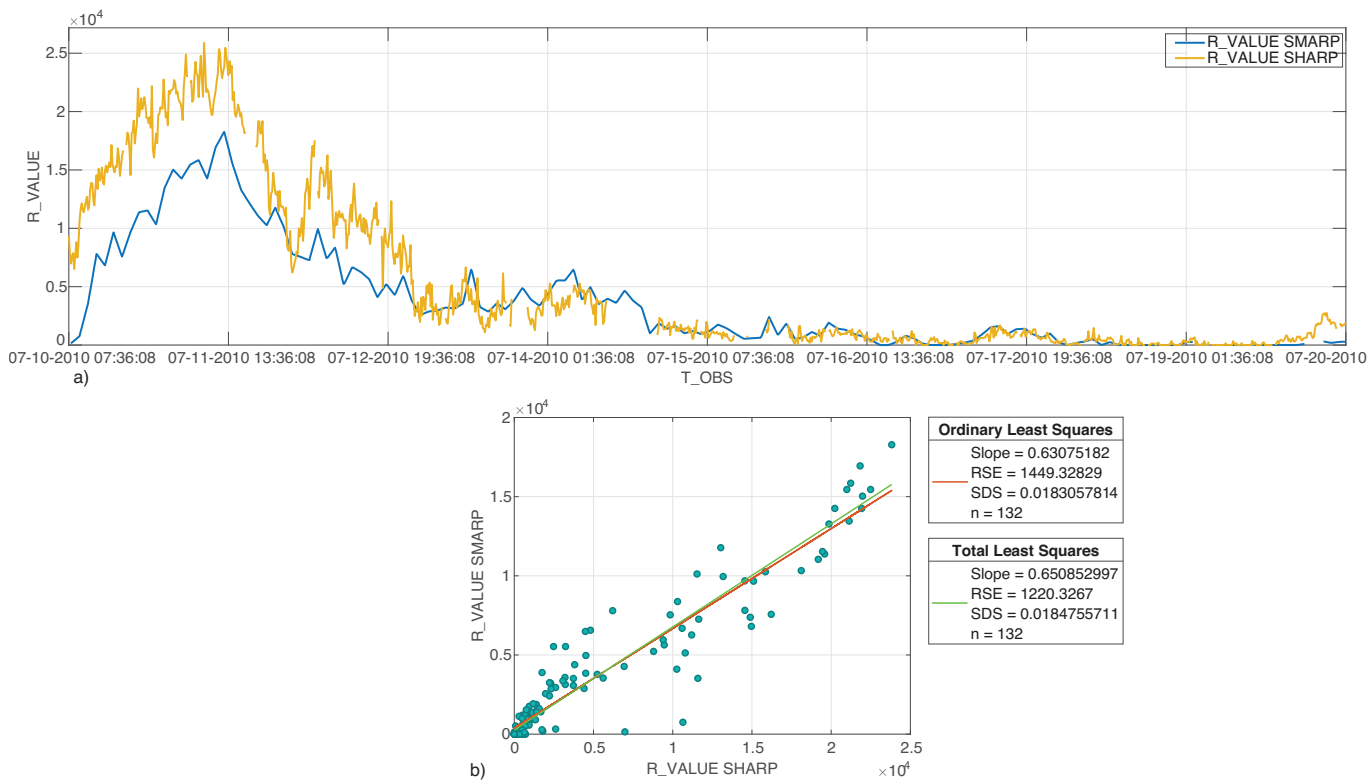

**Figure 2.** The R-value for NOAA active region 11087 plotted as a time series (panel a) and on a scatter diagram (panel b), with the OLS first-order polynomial regression as a solid red line, and TLS as a green line. The slope, RSE and SDS for both regression methods are also shown.

### 2.4. *GOES X-ray Data and the Daily Flare Index*

The daily flare index is given in units of Joule/m², and is calculated from the NOAA Solar and Geophysical Activity Summary data[2], as a sum of the product of GOES X-ray flare magnitude and duration, for all C, M, and X-class flares during a day. Flares corresponding to the Stonyhurst longitudes below -65 and above +65 degrees were excluded from the calculation.

The present study also utilizes 1-minute averaged GOES-15 Soft X-ray data, which are here time-matched with disc-integrated 12-minute SHARP time series, creating a high-cadence data product for the period from September 1, 2010, to March 23, 2017, encompassing the majority of Solar Cycle 24. The data represent the measurement of the Soft X-ray flux in the 1-8 Å wavelength range (only the long channel of GOES is used for the analysis) averaged over one minute of the observations. The original data are available at the National Center for Environmental Information archive[3] by the National Oceanic and Atmospheric Administration (NOAA).

## 3. DATA ANALYSIS

### 3.1. *Time-Lagged Cross-Correlation*

The Pearson correlation coefficient (PCC) (Pearson 1896) alone is not sufficient to establish causality (Wiedermann & Eye 2016, p. 64). However, it is well-known that large-scale flaring ARs are associated with magnetic flux emergence (Toriumi et al. 2017; Toriumi & Wang 2019; Toriumi 2022; van Driel-Gesztelyi & Green 2015). In addition, various SHARP and SMARP parameters, including the total unsigned flux and R-value, are shown to have high predictive capability in various machine learning models (Bobra & Couvidat 2015; Kasapis et al. 2022; Deshmukh et al. 2022a,b; Pham et al. 2019; Sinha et al. 2022). Here, we take a statistical approach.

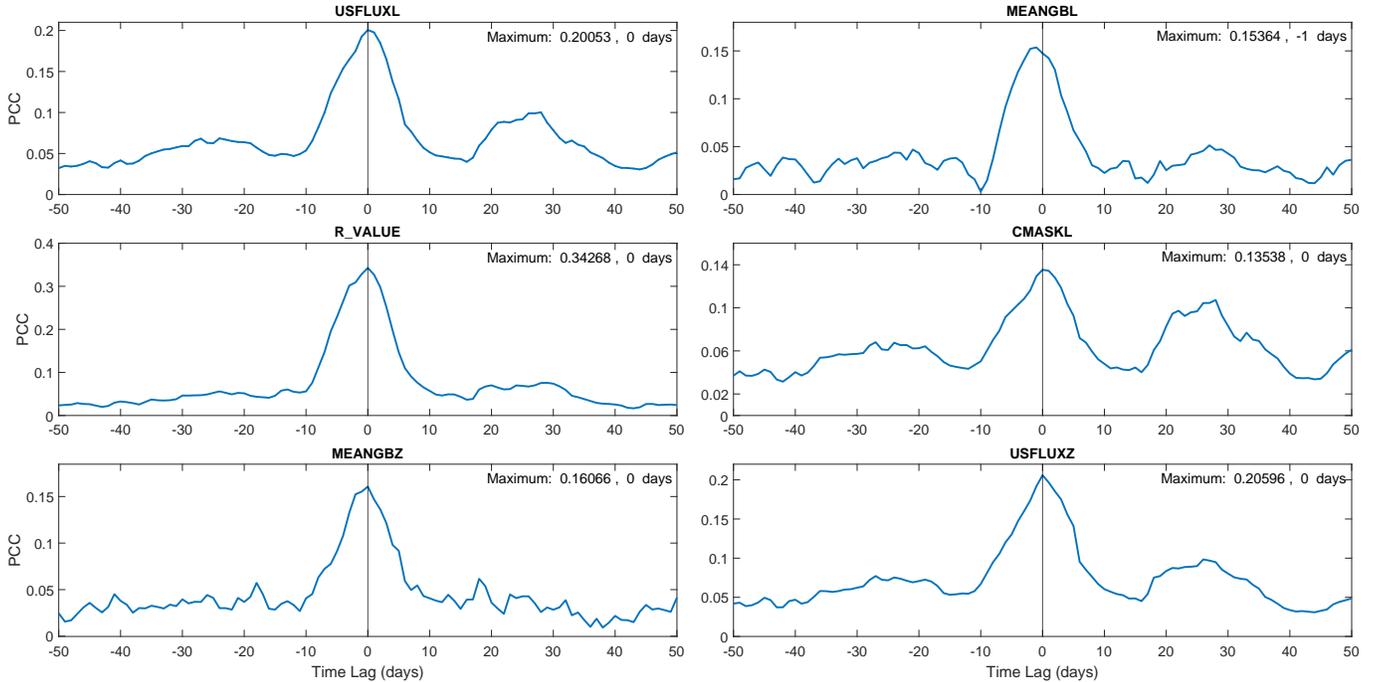

**Figure 3.** Time-Lagged Cross Correlation between selected daily-averaged merged SMARP-SHARP keywords (Table 1) the daily flare index, with the correlation coefficient on the vertical axis, and time lags (representing 1-day offsets) are marked at 10-day intervals on the horizontal axis. The zero offset condition is marked by a vertical line for reference. Left-hand offsets indicate the displacement of the flare data earlier in time with respect to the keyword.

The daily flare index (Section 2.4) is derived from GOES X-ray data, and as such, it is not specific with respect to the location of the source AR or flare. As described in Section 2.2, we also derived an abridged time series based on the merged SHARP-SMARP dataset. This time series is also not AR-specific, rather, it is produced by taking for each observation time the record with the highest R-value. For the purpose of performing cross-correlation analysis, we further reduce the resulting time series by averaging the keywords for each day, in order to achieve the same vector length as that of the daily flare index. The two can then be cross-correlated as follows.

Time-lagged cross-correlation (TLCC) involves calculating the PCC score on time series, which have been offset with respect to one another by a certain number of time lags. If increased amplitude in the SHARP-SMARP parameters consistently precedes increases in flare activity, this would be indicated in the TLCC profile by a non-centered correlation maximum, offset by the average lead time. In this study, 50 positive 1-day time lags, and 50 negative 1-day time lags are introduced between each of the daily-averaged SHARP-SMARP parameters and the daily flare index. For each time lag, the PCC score is computed using Matlab's *corr()* function, which ignores any NaN-containing pairs when the " 'rows','complete' " option is set; therefore, data interpolation is not necessary. Positive lags mean that the flare index was delayed in time relative to a keyword, and negative lags mean that it was ahead in time relative to the keyword, by the number of days indicated on the horizontal axis (Figure 3). This technique forms the basis for our first TLCC analysis.

In the second TLCC analysis, the same technique is applied to time-matched 12-minute SHARP data and GOES-15 1-minute averaged soft X-ray long-wavelength data. In this case, we use logarithmic values in order to compensate for the large discrepancy on the numerical scale between the data from the GOES X-ray detectors and the HMI instrument. The correlation scores are computed for ±1200 12-minute lags, representing a maximum offset of 10 days in each time direction. The purpose here is two-fold: to allow the visualization of the high-correlation portion of the profile in more detail and secondly, to examine the TLCC between two minimally-processed datasets, related to those used in the first analysis. Using a shorter offset frame also allows for a meaningful calculation of the statistical center of the curve (Figure 4).

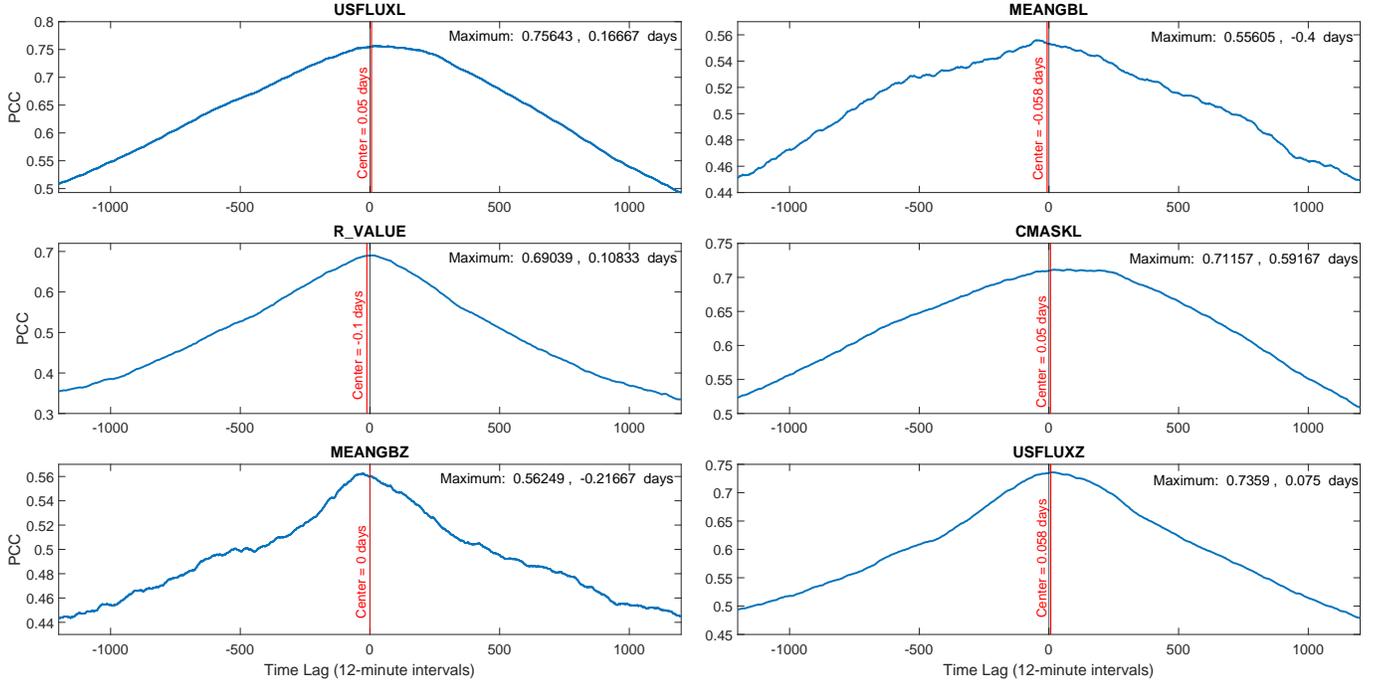

**Figure 4.** Time-lagged cross-correlation between GOES-15 X-ray and HMI SHARP keywords, showing the magnitude and offset of each TLCC maximum (printed in the top-right corner of each subplot) as well as the statistical center of each curve, marked by a vertical red line. The horizontal scale shows the number of offsets in 12-minute lags. Left-hand offsets indicate the displacement of the GOES-15 data earlier in time with respect to the keyword.

### 3.2. *Rolling Window Time-Lagged Cross-Correlation*

While the TLCC method gives the average representation of the time-lagged cross-correlation profile between two signals, it does not show how this interaction varies with time. In order to examine this temporal variation, it is necessary to split the time-matched signal pair into consecutive data segments, or bins, and repeat the TLCC method for each bin. This can be done in one of the following two ways: the first involves performing the TLCC on non-overlapping sequential data bins, whereas the second involves repeating the TLCC analysis by offsetting the data bins incrementally by one data point, which corresponds to the cadence time interval. In the latter case, there will be overlap between consecutive bins: this technique is referred to as the Rolling-Window Time-Lagged Cross-Correlation (RWTLCC). Both of these methods can be used in order to investigate how the leader-follower relationship varies with time – the main difference is that the RWTLCC method allows for a more detailed visualization (Cheong 2022) and, thus, is the method used in the present work. The rolling-window technique is applied to the time-matched daily-averaged SHARP-SMARP data and the daily flare index, which is the same data referred to in Section 3.1. The data are divided into 9630 consecutive one-day bins, and for each bin, the TLCC is calculated with ±50 one-day offsets in each direction, as was done with the TLCC method (Section 3.1). The resulting plot is illustrated in Figure 5.

### 4. DISCUSSION

The parameters used to perform the cross-calibration between the SMARP and SHARP keywords are shown in Table 2. One limitation of the present work is that the rescaling involved using both the slope and the intercept, which resulted in some negative values in R_VALUE and USFLUXZ, although in most cases, the intercept was close to the origin, as expected. It is also possible to force the linear regression to pass through the origin, or to apply a non-linear regression technique, which may be an objective of future work.

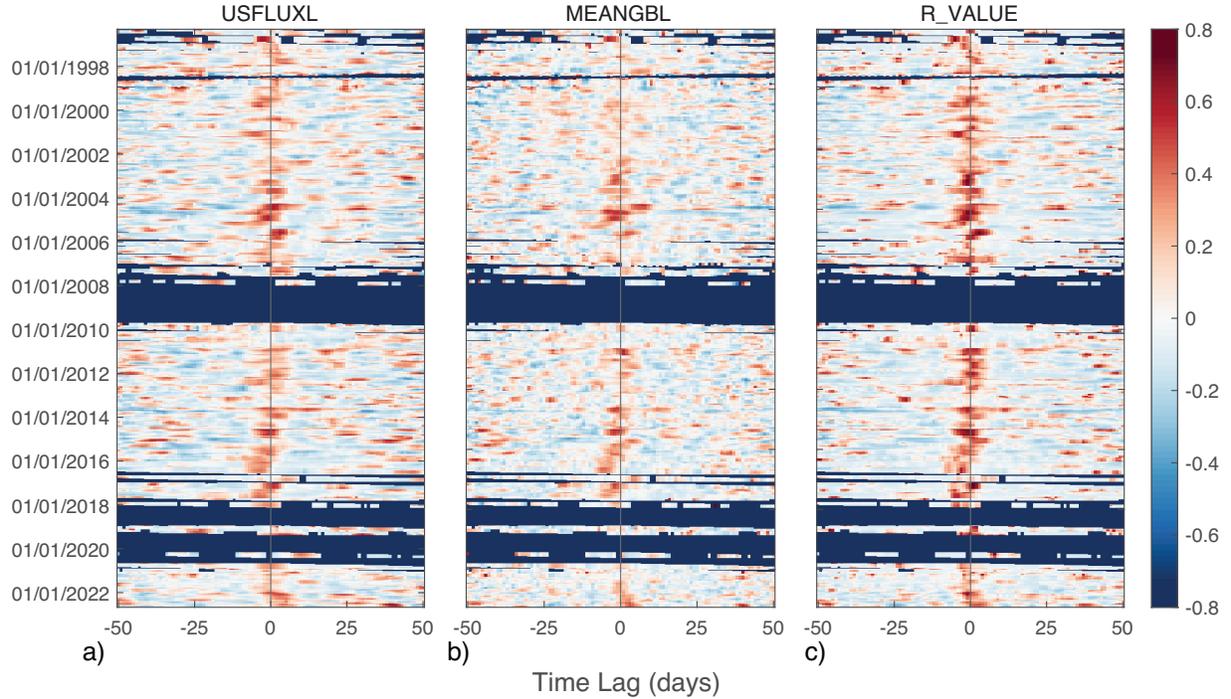

**Figure 5.** Rolling-Window Time-Lagged Cross-Correlation between selected daily-averaged merged SMARP-SHARP parameters: a) USFLUXL (Total line-of-sight unsigned magnetic flux), b) MEANGBL (Mean value of the line-of-sight field gradient), c) R_VALUE (Unsigned flux near polarity inversion lines), and the daily flare index, with the correlation coefficient indicated by the color bar on the right, and time lags on the horizontal axis, at 25-day intervals. The zero offset condition is marked using vertical lines for reference. As in Figure 3, left-hand offsets indicate the displacement of the flare data earlier in time with respect to the keyword.

**Table 2.** Cross-calibration parameters and statistics.

| Keyword | $\mu'$ m | $\mu'$ b | $\mu$ RSE | $\mu$ SDS |
|---------|---------|---------|-----------|-----------|
| (1) | (2) | (3) | (4) | (5) |
| USFLUXL | 1.437776 | -9.34e+20 | 7.85e+20 | 0.082171 |
| MEANGBL | 0.579888 | -0.322213 | 1.712718 | 0.022702 |
| R_VALUE | 0.471202 | 109.1307 | 1641.057 | 0.037700 |
| CMASKL | 0.909633 | -2437.092 | 8120.222 | 0.055187 |
| MEANGBZ | 0.620800 | -4.197899 | 2.447984 | 0.029896 |
| USFLUXZ | 0.733397 | 1.29e+21 | 1.75e+21 | 0.372967 |

NOTE—For each keyword in column 1, the weighted mean slope(2) and weighted mean intercept(3) are given, along with the mean residual standard error(4) and the mean standard deviation of the slope(5) using the TLS method.

### 4.1. *Time-Lagged Cross-Correlation*

TLCC is used in order to establish if a leader-follower relationship exists between two time series, and if so, which time series leads this relationship. In the first TLCC analysis (Figure 3), the peak correlation between the merged SHARP-SMARP keywords and the daily flare index generally occurs at the zero-day offset, which implies that there is no distinct and consistent lead time beyond the 1-day time frame. Notably, a left-hand peak offset and asymmetry are evident with MEANGBL. To a lesser extent, there is also asymmetry with R_VALUE and MEANGBZ, which indicates that a leader-follower dynamic may exist, where these three keywords lead the flare index overall. With the other three parameters (USFLUXL, CMASKL, and USFLUXZ) this peak asymmetry is less pronounced but still indicative of a difference between the lead-up time and the follow-up time. The large secondary peaks seen at around the +27 day mark suggest that many ARs are still showing signs of activity after passing through one Carrington rotation. Overall, the TLSS profiles show increased correlation within ±10 days of the offset reference point, with the peak occurring at, or just prior to it. The parameter exhibiting the highest peak correlation with the flare index is R_VALUE, followed by, in descending order: USFLUXZ, USFLUXL, MEANGBZ, MEANGBL, and CMASKL. There are two reasons for the low overall correlation scores in this first TLCC analysis: scale disparity between the flare index and the keywords, and noise. Log$_{10}$ was not applied, because it would result in imaginary numbers where the keyword values become negative as a consequence of SMARP rescaling. Addressing this limitation is outside the scope of the present work because it is not concerned with different rescaling methods.

One limitation of using one-day averages is the low cadence, making it difficult to identify a prominent peak should it occur within ±1 day from the zero offset reference point. We attempt to address this with the second TLCC analysis, employing 12-minute SHARP data time-matched with GOES-15 1-minute averages. The result (Figure 4) shows much higher overall correlation scores between the keywords and flare activity than was the case in the first TLCC analysis. This is largely due to applying Log$_{10}$ to the data, which was not done prior to the first analysis. Interestingly, the same keywords indicate a lead time in both analyses: MEANGBL, R_VALUE, and MEANGBZ. In Figure 4, only MEANGBL and MEANGBZ show prominent left-hand peak offsets of -0.4 and -0.2 days respectively, and along with R_VALUE, show a left-hand offset of the statistical center of the TLCC curve, although it is not greater than ±0.1 days for any of the six curves. As seen from the first TLCC analysis, and is to be seen from the RWTCLL analysis, this is indeed suggestive of the situation where there is no distinct, immediate, and consistent leader-follower relationship between solar flare activity and the solar magnetic field parameters studied here. This result is largely in agreement with those reported by Whitney Aegerter et al. (2020), Teh (2019), and Teh & Kamarudin (2021).

## 4.2. *Rolling Window Time-Lagged Cross-Correlation*

A predictable offset between two time series is expected to result in a prominent peak in the TLCC profiles, although so far this has not been the case. Therefore, the next step is to investigate how the TLCC profiles vary with time, and whether this variation itself exhibits a discernible pattern. As seen in Figure 5, the three depicted RWTLCC correlation profiles are quite similar, although there are many detailed differences, the analysis of which for all six keywords is beyond the scope of this work. Generally, however, it supports the hypothesis that there are time intervals where the keywords lead the flare index, and others where the flare index leads the keywords. As active regions rotate onto the observed solar disc, some of them may already be erupting, while others may transit it, and erupt on the western limb. Notably, there are many instances in Figure 5, when the peak correlation is offset from zero by ±10 days, which roughly corresponds to the AR transit time. This is consistent with the first TLCC analysis (Figure 3). However, there are many instances, where the peak correlation score is substantially greater in Figure 5 than it is in Figure 3, as expected.

Time intervals containing high peak-correlation values are perhaps the most intriguing aspect of the RWTLCC analysis so far, and it can be hypothesized that they may correspond to peak flare activity. As already mentioned, there are many similarities between the RWTLCC plots of the keywords in Figure 5 (as well as the three not shown) therefore, their average may serve as a representative proxy for a preliminary test of this hypothesis. Computing the arithmetic mean of the six RWTLCC matrices and plotting them opposite of the daily flare index results in Figure 6, which suggests that time intervals with higher correlation peaks generally correspond to increases in C-X class solar flare activity, regardless of the exact location of these peaks with respect to zero.

## 5. CONCLUSIONS

The most significant aspects of this work include rescaling and merging selected SHARP and SMARP magnetic field parameters ("keywords") using the TLS regression technique, and performing time-lagged cross-correlation in order to investigate the lead time between these parameters and solar flare activity. A consistent offset in the TLCC and RWTLCC profiles would show if there is an optimal flare forecasting window and if any keywords serve as potential flare precursors. One conclusion from this study is that in some cases, the keyword leads the flare, but sometimes the opposite is true. One possible explanation for this is that some active regions continue to erupt after the initial event. Another conclusion is that some keywords may serve as better precursors, for example, MEANGBL, whose offset TLCC peak can be accounted for by noise to some extent, but the strong peak asymmetry cannot. Furthermore, the RWTLCC results for the six keywords lead to the hypothesis that areas of high correlation generally correspond to increased flare activity during the peak solar cycle.

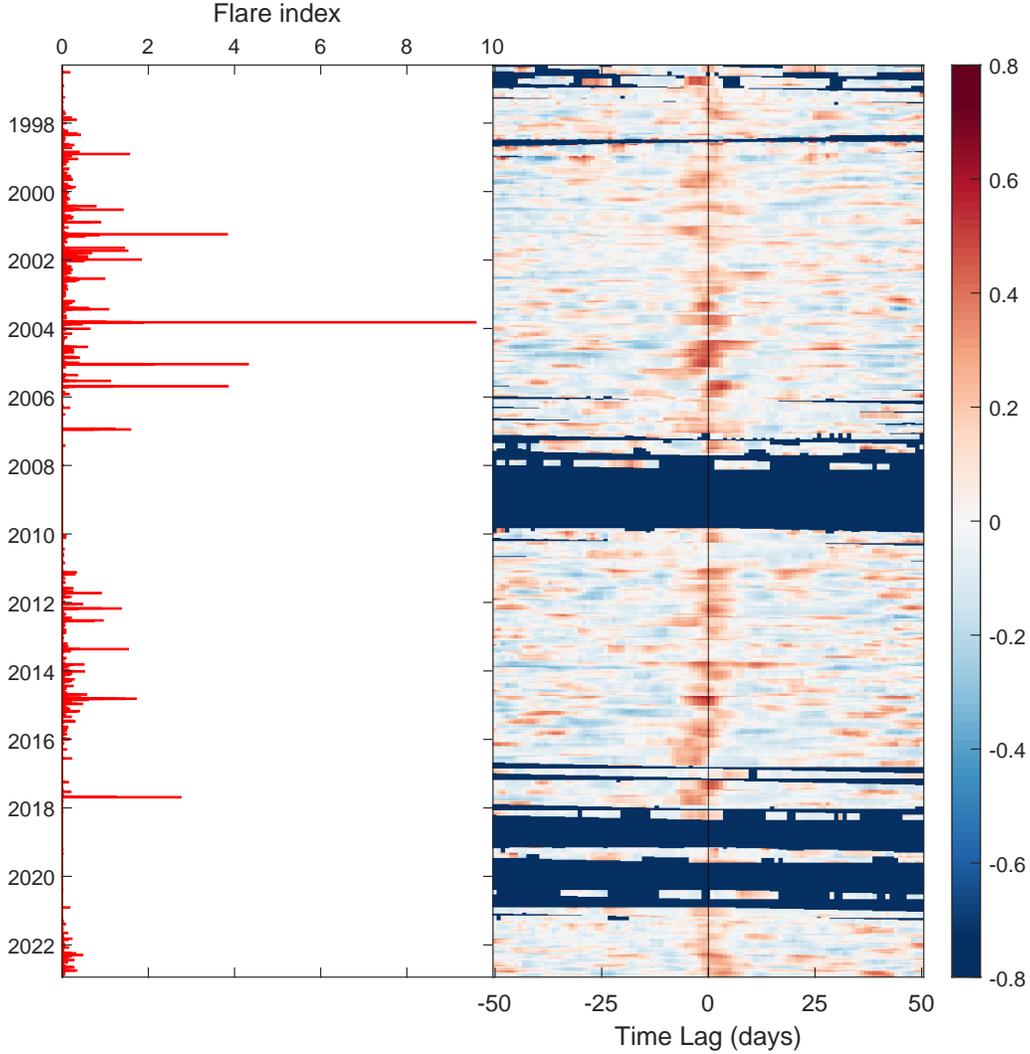

**Figure 6.** The daily flare index (left) and the average of the RWTLCC profiles of the six keywords in Table 2. The correlation coefficient is indicated by the color bar on the right, and time lags are labeled on the lower right-hand scale, at 25-day intervals. As in Figure 5, the zero offset condition is marked using a vertical line for reference, and left-hand offsets indicate the displacement of the flare data earlier in time with respect to the keyword.

One apparent limitation of the flare data is the reliance on only C-X class flares. This may distort the cross-correlation profiles to some extent and introduce large gaps, especially during solar minima. In general, there are flares during solar "quiet" periods; however, the problem is exacerbated when applying the longitude truncation at ±65°, which results in the exclusion of approximately 7% of C-X class flares from the original dataset. There is some discrepancy in the literature regarding the size of the longitude window corresponding to the acceptable HMI signal-to-noise ratio: it is reportedly ±70° from the central meridian in Bobra & Couvidat (2015) and Abduallah et al. (2023), although Lee et al. (2020), and Yi et al. (2021) use ±60°.

Different studies use different criteria for space weather forecasting and for parameter ("feature") ranking. In the present study, no attempt is made to predict space weather; however, we demonstrate that the mean value of the line-of-sight field gradient (MEANGBL) and the mean value of the vertical field gradient (MEANGBZ) may have some value in forecasting C-X class flares. Most machine learning studies exclude MEANGBZ, because it received a low Fisher score in Bobra & Couvidat (2015), though it must be kept in mind, that in that particular work, as well as many others, feature ranking is based on a 24-hour prediction window. It is perhaps unsurprising, that other workers also reported relatively low scores for MEANGBZ (Liu et al. 2019; Wang et al. 2019; Zhang et al. 2022), while MEANGBL and CMASKL are rarely mentioned at all.

In many current machine learning models, the best-performing predictors are, in no particular order: total unsigned current helicity (TOTUSJH), total unsigned vertical current (TOTUSJZ), total magnitude of Lorentz force (TOTBSQ), total photospheric magnetic free energy density (TOTPOT), sum of the modulus of the net current per polarity (SAVNCPP), absolute value of the net current helicity (ABSNJZH), area of strong field pixels in the active region (AREA_ACR), total unsigned flux (USFLUX), and R_VALUE (Jonas et al. 2018; Ma et al. 2017; Hamdi et al. 2017; Chen et al. 2019; Lee et al. 2020; Lim et al. 2019a,b; Liu et al. 2017; Zheng et al. 2023; Zhang et al. 2022; Vysakh & Mayank 2023). To date, more than 50 studies utilizing SHARP keywords have been published. Many of them focus primarily on the development and performance of different machine learning algorithms and tend to overlook SMARP data. They also tend to consider only a limited number of magnetic field parameters, which is also a limitation in the present study. Nevertheless, the merged SMARP-SHARP data series are shown to have potential for use in machine learning applications for forecasting space weather events, including SEP events (Kasapis et al. 2024, accepted).

Future work may involve expanding the merged SMARP-SHARP dataset by including additional keywords and adding B-class flares to the daily flare index. In the second TLCC analysis, it is possible to reduce the 12-minute SHARP data by choosing time slices corresponding to peak R_VALUE. The RWTLCC analysis may be modified by using the combined SHARP-SMARP keywords downsampled to a uniform 96-minute cadence, and time-matching them with a 96-minute averaged GOES X-ray flux, which would still allow for a greater level of detail than does the 1-day flare index. One limitation of the GOES X-ray flux, however, is that it is not coordinate-specific, and therefore difficult to match with any particular active region. Lastly, an RWTLCC analysis between SHARP and GOES-15 X-ray time series may test the hypothesis, that areas of high correlation generally correspond to increased flare activity during the peak solar cycle.

## ACKNOWLEDGMENTS

We thank Joel Tellinghuisen at Vanderbilt University (Department of Chemistry) for his valuable comments on the mathematical formulation of TLS regression parameters. This research was supported by NASA Early Stage Innovation program grant 80NSSC20K0302, NASA LWS grant 80NSSC19K0068, NSF EarthCube grants 1639683, 1743321 and 1927578, and NSF grant 1835958. VMS acknowledges the NSF FDSS grant 1936361 and NASA COFFIES DSC grant.

*Software:* Matlab 2022a